\documentclass{article}

\usepackage[preprint]{neurips_2026}

\usepackage[utf8]{inputenc} % allow utf-8 input
\usepackage[T1]{fontenc}    % use 8-bit T1 fonts
\usepackage{hyperref}       % hyperlinks
\usepackage{url}            % simple URL typesetting
\usepackage{booktabs}       % professional-quality tables
\usepackage{amsfonts}       % blackboard math symbols
\usepackage{nicefrac}       % compact symbols for 1/2, etc.
\usepackage{microtype}      % microtypography

\usepackage[ruled,vlined,linesnumbered]{algorithm2e}
\usepackage{amsmath}
\usepackage{enumitem}
\usepackage{graphicx}
\usepackage{subcaption}
\usepackage{tabularx}
\usepackage{xcolor}

\title{AuditoryHuM: Auditory Scene Label Generation and Clustering using Human-MLLM Collaboration}

\author{
  {Henry Zhong} \quad {Jörg M. Buchholz} \\
  Macquarie University \\
  Sydney, Australia \\
  \texttt{\{henry.zhong, jorg.buchholz\}@mq.edu.au} \\
  \And
  {Julian Maclaren} \quad {Simon Carlile} \quad {Richard F. Lyon} \\
  Google Research Australia \\
  Sydney, Australia \\
  \texttt{jmaclaren@xwf.google.com, \{scarlile, dicklyon\}@google.com} \\
}

\begin{document}

\maketitle

\begin{abstract}
Manual annotation and clustering of audio datasets is labour intensive. We introduce AuditoryHuM, a training-free framework for the unsupervised discovery and clustering of auditory scene labels using human-Multimodal Large Language Model (MLLM) collaboration. Leveraging MLLMs, our framework generates contextually relevant labels for audio data. To ensure label quality and mitigate hallucinations, zero-shot learning (Human-CLAP) quantifies the alignment between generated text labels and raw audio. A targeted human-in-the-loop intervention, refines only the lowest aligned pairs. The discovered labels form an interpretable alignment vector to group audio into cohesive clusters. The framework was evaluated across three auditory scene datasets (ADVANCE, AHEAD-DS, and TAU 2019), achieving a 96.1\% reduction in human labour for ADVANCE during testing. Downstream models trained on our clusters exhibit enhanced classification accuracy vs the baseline (0.79 to 0.84) in clean acoustic environments, though performance scales down in dense, complex soundscapes. The project page: \url{https://github.com/Australian-Future-Hearing-Initiative}
%TODO
\end{abstract}

\section{Introduction}\label{sec:intro}
Auditory scene recognition is a core feature of devices such as hearing aids \cite{vivek2020acoustic} and smart home assistants \cite{haeb2019speech}. Scene recognition enables devices to improve users’ experience by triggering specific signal processing routines \cite{cauchi2018hardware} in response to the soundscape, such as wind noise reduction, beam forming, music mode, etc. A method of auditory scene recognition is to use deep learning (DL) models \cite{zaman2023survey} trained on labelled audio datasets.

Creators of audio datasets generally define a set of labels and assign labels by listening to each recording. However, manual annotation is difficult to scale as it is labour intensive \cite{moore2017neural}. This paper proposes an alternative solution: instead of fitting the data to predefined labels, labels are fit to the data. This can be achieved using Multimodal Large Language Models (MLLMs) \cite{yin2024survey}, which enable the automated discovery of labels tailored specifically to the audio. However, to mitigate the risk of MLLM hallucinations \cite{bai2024hallucination} a method for measuring label quality is necessary.

Zero-shot learning techniques \cite{wu2023large} have enabled the scoring of alignment between labels and audio data. These scores allow for a quantitative evaluation of label relevance, ensuring MLLM generated labels are grounded in the actual audio content. This metric driven framework can be further enhanced by a human-in-the-loop (HITL) approach. Where manual review is targeted at the least aligned audio label pairs to improve labelling quality and ensure labels align with actual human perception.

Acoustically related samples can then be clustered using a label alignment vector. This is a humanly interpretable description of acoustic characteristics of each sample, constructed from the top scoring labels of each sample. While clustering algorithms effectively cluster data, the contents of each cluster cannot be determined at a glance. A label distribution vector can be derived from each cluster to generate a concise, semantic composite label for the entire cluster. This transition from sample specific labels to a standardised taxonomy enables the downstream training of auditory scene recognition models.

The contributions of this paper are the following:

\begin{itemize}
\item We introduce AuditoryHuM, a novel training-free framework designed for unsupervised auditory scene label discovery and clustering using human-MLLM collaboration.
\item We validate label alignment using zero-shot learning and targeted HITL intervention, governed by diminishing returns optimisation constraints, to mitigate hallucinations and ensure labels are grounded in human perception. Final scene clustering is based on a human interpretable label alignment vector and composite labels are generated to describe each cluster using a label distribution vector.
\item We evaluate AuditoryHuM across multiple auditory scene datasets using a suite of metrics. This includes our own novel ones ($\mu_c$ and $\mu_{x\%}$) to assess the alignment of labels, the coherence of clusters, and the learnability of generated clusters for downstream model training.
\end{itemize}

\section{Related Work}\label{sec:related}
Auditory scenes can be categorised through unsupervised learning. Older approaches rely on handcrafted features, such as mel frequency cepstral coefficients (MFCC) \cite{sebastian2025audio}, while DL has extended these capabilities using deep neural networks. For example, Fiorio et al. \cite{fiorio2025unsupervised} utilised variational autoencoders, while Hershey et al. clustered audio embeddings \cite{hershey2016deep}. AuditoryHuM takes an additional intermediate step of generating text labels using MLLMs. This allows thematically similar sounds to be grouped based on text labels, which are more humanly interpretable than audio embeddings.

A number of MLLMs \cite{team2025gemma}\cite{chu2024qwen2}\cite{xu2025qwen2}\cite{xu2025qwen3omnitechnicalreport}\cite{qwenteam2026qwen35omnitechnicalreport}\cite{goel2025audio} are able to incorporate both audio and text as input. MLLMs convert audio into audio tokens which exist in the same embedding space as text tokens. AuditoryHuM uses MLLMs to analyse and provide descriptive labels for auditory scenes.

The alignment of text labels and audio data can be calculated using Contrastive Language Audio Pretraining (CLAP) models \cite{wu2023large}\cite{takano2025human}\cite{yuan2024t}\cite{li2024advancing}. AuditoryHuM compares the similarity of CLAP audio and text embeddings to quantitatively evaluate the alignment of labels produced by MLLMs, in order to mitigate the effect of hallucinations.

Prior audio captioning frameworks generate free form text descriptions by leveraging metadata driven LLMs (\emph{WavCaps} \cite{mei2024wavcaps}), audio visual models (\emph{Sound-VECaps} \cite{yuan2025sound}, \emph{Auto-ACD} \cite{sun2024auto}), or direct audio architectures (\emph{AudioSetCaps} \cite{bai2025audiosetcaps}). AuditoryHuM differs by incorporating an optimized HITL refinement stage and grouping audio via semantic alignment vectors. Unlike unstructured free form captions, bounded, standardised clusters are practical for downstream acoustic model training.

\section{Method}\label{sec:method}
\begin{figure}
\centering
\captionsetup{justification=centering}
\includegraphics[width=0.5\textwidth]{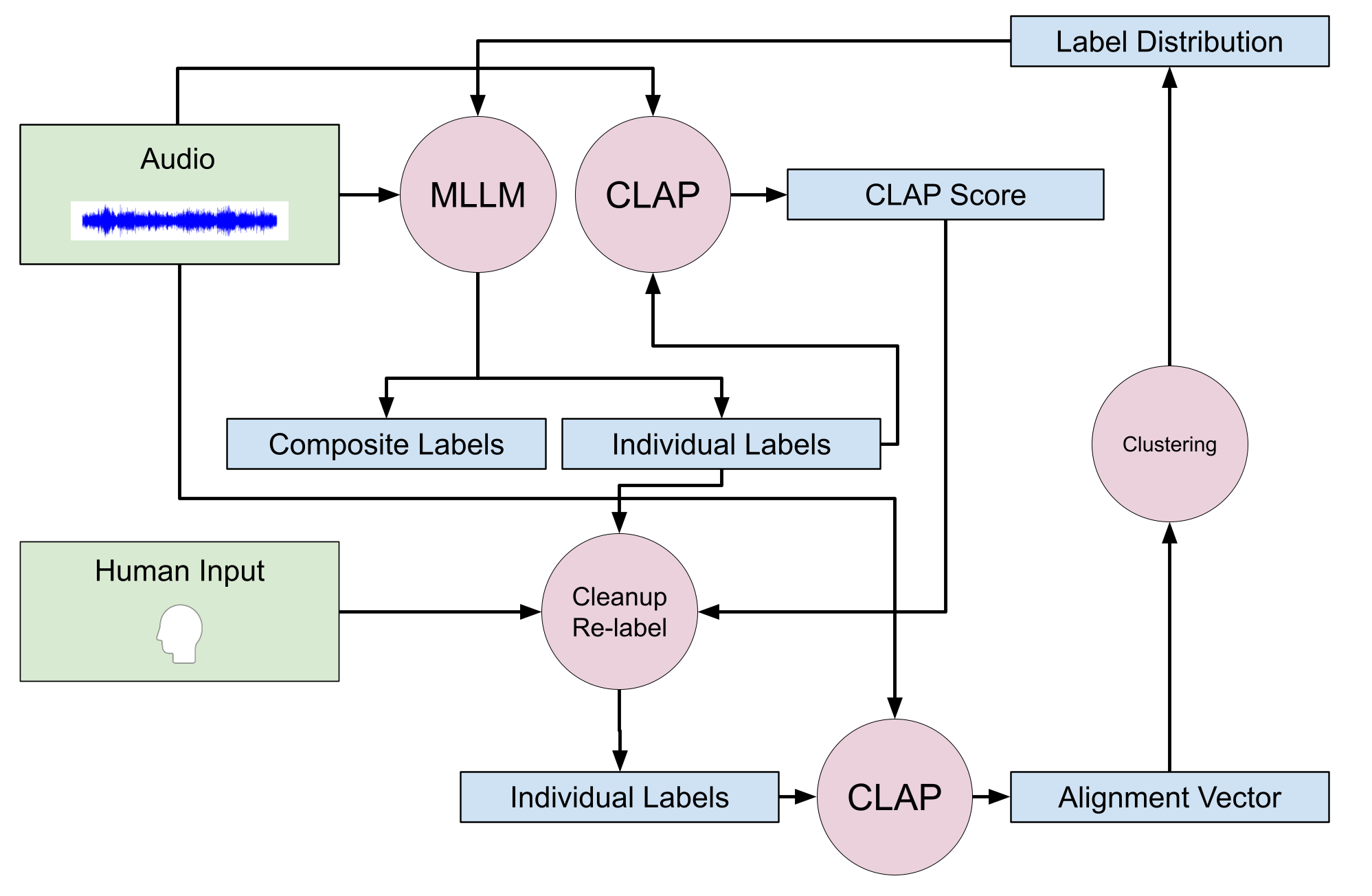}
\caption{The dataflow diagram of the AuditoryHuM processes, showing the steps from MLLM label discovery, human-in-the-loop refinement, cleanup, alignment, and clustering.}
\label{fig:cluster_arch}
\end{figure}

The AuditoryHuM workflow is outlined in Figure \ref{fig:cluster_arch}. Initial labels are generated for all audio samples using a MLLM with the prompts detailed in Table \ref{tab:prompts}. The starting prompt elicits short, descriptive word pairs rather than long sentences to ensure interpretability.

Labels undergo a preprocessing cleanup step: text is converted to lowercase, non-alphanumeric characters, punctuation, whitespace, and non-printing characters are removed, and sentences are truncated to the first two words. This pipeline cleans up hallucinated long sentences or non-English labels that disrupt interpretability.

Audio label alignment is quantified via cosine similarity using a CLAP model, yielding a \emph{CLAP score}. This metric enables a targeted HITL approach, directing human annotators to relabel the least aligned samples, as opposed to relabelling based on opinion or randomly. The highest scoring label per sample is retained. The mean of these maximum scores, denoted $\mu_c$, serves as the core evaluation metric for MLLM performance. To track alignment improvements within the worst performing cohorts before and after human intervention, the conditional mean $\mu_{x\%}$ of the lowest $x\text{th}$ percentile is used. Definitions are as follows.
\begin{align*}
    A     &: \text{Audio dataset}                                & \hat{c}_{a} &= \max_{l \in L_a} c(a, l) \\
    L_a   &: \text{Set of labels for sample } a                  & \mu_c       &= \frac{1}{|A|} \sum_{a \in A} \hat{c}_a \\
    a     &\in A, l \in L_a                                      & P_x         &: \text{$x$-th percentile of all scores $\{\hat{c}_a\}$} \\
    c(a, l) &: \text{CLAP score between $a$ and $l$ }            & \mu_{x\%}   &= \frac{1}{| \{ a \in A \mid \hat{c}_a \le P_x \} |} \sum_{a \in A, \hat{c}_a \le P_x} \hat{c}_a
\end{align*}
Following label generation for each audio sample, the best scoring label for every sample $l_a^*$ is collected together and form the ordered label set $L_A^*$. The \emph{CLAP score} of every audio sample and every label from $L_A^*$ are computed, resulting in an alignment vector $v_a$ per sample. This vector leverages cross sample label alignment similarities to describe audio characteristics in a human interpretable feature space, and is used to group audio using a clustering algorithm. Once clusters are finalised, the label distribution vector $F_i$ is fed into a MLLM to generate a final composite label so the contents of $ith$ cluster can be determined at a glance. More definitions are as follows.
\begin{align*}
    l_a^*   &= \arg\max_{l \in L_a} c(a, l)                                      & A_i       &\subseteq A \\
    L_A^*   &= \{l_1, l_2, \dots, l_M\} = \{l_a^* \mid a \in A\}                 & F_i       &= [f_{(i,1)}, f_{(i,2)}, \dots, f_{(i,M)}] \\
    v_a     &= \left[ c(a, l_j) \right]_{j=1}^M, \quad \text{where } l_j \in L_A^* & f_{(i,j)} &= \sum_{a \in A_i} \mathbb{I}(l_a^* = l_j)
\end{align*}

\subsection{Cluster Count}\label{sec:cluster_count}
Selection of the cluster count $k$ relies on the silhouette score $s(k)$ \cite{rousseeuw1987silhouettes}, targeting the point $k^*$ where the score stabilises. Discovery of $k^*$ involves evaluating $s(k)$ from $k=2$ to $k_{max}$ using a moving variance $Var(k, w)$ with window size $w$ to detect stability. Within this range, the minimum and maximum variances are denoted $Var_{min}$ and $Var_{max}$. The algorithm increments $k$ from $2$ to $k_{max} - w$ and terminates at the first instance where the normalised variance drops such that $Var(k, w) / (Var_{max} - Var_{min}) \le t$, where the threshold $t$ controls cluster granularity. If $Var_{min} = Var_{max}$, clustering is aborted. The logic is detailed in Algorithm \ref{alg:cluster_selection}.

\subsection{Human Labelling Strategy}\label{sec:human_labelling_strategy}
The human labelling strategy is designed to maximise the efficacy of manual annotation by targeting samples with the lowest \emph{CLAP score} and terminate when either of the following criteria are met.

\textbf{Resource Boundary:} $R < q \times (n+1)$ Where $R$ is the total allocated human labour budget, $q$ denotes the time required for a single annotation, and $n$ is the current cumulative count of human-labelled samples. \\
\textbf{Diminishing Returns Constraint:} $\frac{1}{m}\sum_{i=n-m}^{n} \Delta c_i \le s$ Where $\Delta c$ is the alignment gain between MLLM and human label scores. $m$ is the moving average window length, and $s$ is the termination threshold.

The first condition establishes an upper bound based on resource depletion. The second condition identifies the convergence point where the moving average of alignment improvements drops below $s$, indicating that additional manual labelling yields negligible marginal utility.

\section{Results}\label{sec:results}
The three tested datasets were: AuDio Visual Aerial sceNe reCognition datasEt (ADVANCE) \cite{hu2020cross}, Another HEaring AiD scenes Data Set (AHEAD-DS) \cite{zhong2026datasetmodelauditoryscene}, and TAU Urban Acoustic Scenes 2019 Development dataset (TAU 2019) \cite{heittola_toni_2019_2589280}. The datasets represent a broad range of auditory scenes including in-vehicle, urban, suburban, rural, and natural environments. Unless otherwise stated the following configurations were used for all tests. The MLLM was Qwen 2.5 Omni 3B \cite{xu2025qwen2}. The CLAP implementation was Human-CLAP \cite{takano2025human}. The clustering algorithm was spectral clustering \cite{von2007tutorial}.

\subsection{Label Alignment}\label{sec:label_alignment}
This study compares three MLLMs: Gemma 3N E2B \cite{team2025gemma}, Qwen 2 Audio 7B \cite{chu2024qwen2}, and Qwen 2.5 Omni 3B. These represent lightweight, established , and near state-of-the-art (SOTA) architectures, respectively. The $\mu_c$ values for each MLLM and dataset pair are shown in Table \ref{tab:clap_scores}. $\mu_c$ represents the mean of the highest \emph{CLAP score} for each sample. $\mu_c$ ranges from 1 to $-1$, a value of 1 is perfect alignment and $-1$ is perfect misalignment. A higher value is better. The results indicated the most sophisticated model, Qwen 2.5 Omni 3B, produced the best aligned labels.

The $\mu_{1\%}$ values for each MLLM  and dataset pair before and after human relabelling are shown in Table \ref{tab:clap_percentile}. $\mu_{1\%}$ represents the mean of the $1st$ percentile (bottom $1\%$) of \emph{CLAP scores}. $\mu_{1\%}$ also ranges from 1 to $-1$, a value of 1 represents perfect alignment and $-1$ is perfect misalignment. A higher value is better. During relabelling, a human labeller was tasked with identifying the most prominent sound in each sample. The human provided labels significantly boosted alignment of the mean of the bottom $1\%$ of \emph{CLAP scores}. This highlighted the efficacy of strategically targeted human intervention. Human provided labels still scored lower than $\mu_c$. This was a sign that the bottom $1\%$ were challenging samples to categorise. A number of challenging audio samples were recorded in very noisy, very quiet, intentionally mixed, or complicated multilayered soundscapes, where it was difficult to find a single well-aligned label.
\begin{table*}
    \centering
    \caption{The $\mu_c$ values for each dataset and MLLM pair, rounded to two decimal places. $\mu_c$ ranges from 1 to $-1$, a value of 1 represents perfect alignment and $-1$ is perfect misalignment. A higher value is better.}
    \label{tab:clap_scores}
    \begin{tabularx}{\textwidth}{p{0.15\textwidth}p{0.2\textwidth}p{0.2\textwidth}p{0.2\textwidth}}
    \hline
        $\mu_c$ & Gemma 3N E2B & Qwen 2 Audio 7B   & Qwen 2.5 Omni 3B  \\
    \hline
        ADVANCE  & 0.51 & 0.56 & 0.61 \\
        AHEAD-DS & 0.42 & 0.51 & 0.55 \\
        TAU 2019 & 0.48 & 0.48 & 0.49 \\
    \hline
    \end{tabularx}
\end{table*}

\begin{table*}
    \centering
    \caption{The $\mu_{1\%}$ values for each dataset and MLLM pair before and after relabelling by a human, rounded to two decimal places.  $\mu_{1\%}$ represents the mean of the $1st$ percentile (bottom $1\%$) of \emph{CLAP scores}.  $\mu_{1\%}$ ranges from 1 to $-1$, a value of 1 represents perfect alignment and $-1$ is perfect misalignment. A higher value is better.}
    \label{tab:clap_percentile}
    \begin{tabularx}{\textwidth}{p{0.15\textwidth}p{0.2\textwidth}p{0.2\textwidth}p{0.2\textwidth}}
        \hline
            $\mu_{1\%}$ & Gemma 3N E2B & Qwen 2 Audio 7B   & Qwen 2.5 Omni 3B  \\
        \hline
            ADVANCE  & -0.06 & -0.14 & 0.13 \\
            AHEAD-DS & -0.19 & -0.26 & 0.07 \\
            TAU 2019 & 0.07  & 0.00  & 0.10 \\
        \hline
            ~        & Human Labelled  & Human Labelled & Human Labelled \\
        \hline
            ADVANCE  & 0.30 & 0.31 & 0.30 \\
            AHEAD-DS & 0.48 & 0.48 & 0.22 \\
            TAU 2019 & 0.13 & 0.30 & 0.25 \\
        \hline
    \end{tabularx}
\end{table*}

\subsection{Label Alignment Comparison with Other Methods}\label{sec:label_alignment_comparison_with_other_methods}
The comparison of $\mu_c$ using several methods is shown in Table \ref{tab:alignment_compare}. AuditoryHuM produced the best results. The zero-shot approach was the baseline which utilised the labels provided with each dataset. The provided labels led to lower alignment as ADVANCE and TAU 2019 labels were based on recording location rather than audio content. AHEAD-DS had the best provided labels, due to being based on audio content, but was weakly labelled. The same MLLM as used in AuditoryHuM (Qwen 2.5 Omni 3B) generated captions using the prompt recorded in Table \ref{tab:prompts}. Captioning also produced lower alignment, as free form captions contain structural filler. This highlights that fitting the labels to audio content of each sample leads to best alignment.
\begin{table*}
    \centering
    \caption{$\mu_c$ represents mean alignment of labels produced by AuditoryHum vs zero-shot using provided labels vs captioning. A higher value is better.}
    \label{tab:alignment_compare}
    \begin{tabularx}{\textwidth}{p{0.15\textwidth}p{0.2\textwidth}p{0.2\textwidth}p{0.2\textwidth}}
    \hline
        $\mu_c$ & AuditoryHuM & Zero-Shot   & Captioning  \\
    \hline
        ADVANCE  & 0.61 & 0.25 & 0.42 \\
        AHEAD-DS & 0.55 & 0.40 & 0.46 \\
        TAU 2019 & 0.49 & 0.33 & 0.19 \\
    \hline
    \end{tabularx}
\end{table*}

\subsection{CLAP Implementation Comparison}\label{sec:clap_implementation_comparison}
The comparison results for CLAP implementations are summarised in Table \ref{tab:clap_imp}. For this analysis, the lowest $1\%$ of aligned labels, as identified by Human-CLAP, were reprocessed using LAION-CLAP \cite{wu2023large}. While Human-CLAP demonstrated that human-provided labels align better with audio content, LAION-CLAP rated human and MLLM-generated labels as having similar alignment. This observation reflects the bias in the training data, which includes significant webscraped machine generated labels for LAION-CLAP, while Human-CLAP was finetuned using human labelled data. Given that the objective was to determine if MLLM labels could replace manual annotation, this discrepancy indicates that LAION-CLAP does not sufficiently reflect human perception, rendering it unsuitable for this evaluation.
\begin{table*}
    \centering
    \caption{The $\mu_{1\%}$ values comparing the results from Human-CLAP and LAION-CLAP rounded to 2 decimal places. A higher value indicates better alignment between label and audio content.}
    \label{tab:clap_imp}
    \begin{tabularx}{\textwidth}{p{0.15\textwidth}p{0.35\textwidth}p{0.35\textwidth}}
        \hline
            $\mu_{1\%}$ & Qwen 2.5 Omni 3B + Human-CLAP & Qwen 2.5 Omni 3B + LAION-CLAP \\
        \hline
            ADVANCE  & 0.13 & 0.07 \\
            AHEAD-DS & 0.07 & 0.07 \\
            TAU 2019 & 0.10 & 0.13 \\
        \hline
            ~        & Human Labelled + Human-CLAP  & Human Labelled + LAION-CLAP \\
        \hline
            ADVANCE  & 0.30 & 0.10 \\
            AHEAD-DS & 0.22 & 0.04 \\
            TAU 2019 & 0.25 & 0.11 \\
        \hline
    \end{tabularx}
\end{table*}

\subsection{Label Cleanup}\label{sec:label_cleanup}
The results of label cleanup are summarised in Table \ref{tab:cleanup}. Minimal cleanup only removed whitespace and non-printing characters as required for correct code execution. While default cleanup added lowercasing, removing non-alphanumeric characters and punctuation, and truncating sentences to the first two words. While longer sentences slightly inflated mean CLAP scores, non-English labels had a divergent impact: marginally increasing scores for AHEAD-DS while decreasing them for TAU 2019. Ultimately, the primary reason to shorten labels and remove non-English labels was to increase label interpretability.
\begin{table*}
    \centering
    \caption{The mean \emph{CLAP score} values comparing the results with default or minimal label cleanup.}
    \label{tab:cleanup}
    \begin{tabularx}{\textwidth}{p{0.3\textwidth}p{0.1\textwidth}p{0.25\textwidth}p{0.25\textwidth}}
        \hline
            ~ & Instances & Mean \emph{CLAP score} Minimal Cleanup & Mean \emph{CLAP score} Default Cleanup \\
        \hline
            ADVANCE Long Labels & 578 & 0.60 & 0.57 \\
            ADVANCE Non-English Lab. & 0 & ~ & ~ \\
            AHEAD-DS Long Labels & 781 & 0.60 & 0.58 \\
            AHEAD-DS Non-Eng. Lab. & 4 & 0.14 & -0.07 \\
            TAU 2019 Long Labels & 2726 & 0.50 & 0.49 \\
            TAU 2019 Non-English Lab. & 2 & 0.13 & 0.18 \\
        \hline
    \end{tabularx}
\end{table*}

\subsection{Clustering Comparison with Other Methods}\label{sec:clustering_comparison_with_other_methods}
\begin{figure}
\centering
\captionsetup{justification=centering}
\includegraphics[width=0.65\textwidth]{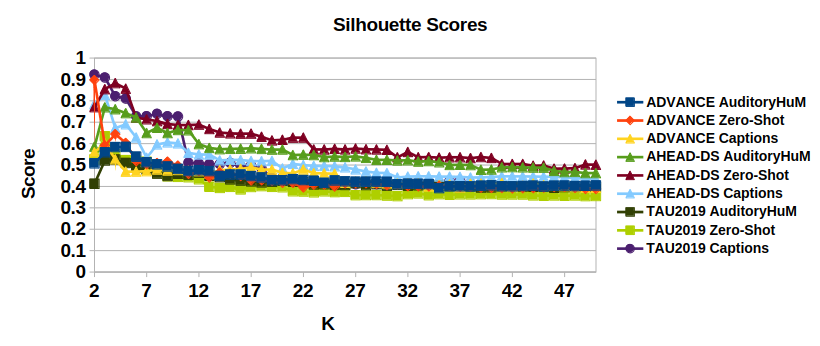}
\caption{Silhouette scores comparing AuditoryHuM vs zero-shot CLAP audio embeddings vs caption sentence transformer embeddings. A higher value is better.}
\label{fig:cluster_compare}
\end{figure}

The comparison of clustering using several methods is shown in Figure \ref{fig:cluster_compare}. AuditoryHuM clustered the alignment vector, the zero-shot approach was the baseline which clustered Human-CLAP audio embeddings, and captions approach clustered sentence transformer \cite{reimers2019sentence} (all-mpnet-base-v2 \cite{all-mpnet-base-v2}) embeddings generated from the MLLM produced captions. The silhouette scores for all methods behave similarly with variance decreasing with increasing $k$. AuditoryHuM offers a distinct advantage, as label aligned vectors remain human interpretable, while alternative approaches rely on opaque embeddings.

The optimal number of clusters ($k^*$) was chosen at the point where the normalised moving variance dropped below half, so the heuristic parameters were configured as follows: $k_{max}=50$, $w=20$, and $t=0.5$. Based on this criteria, the optimal cluster count was determined to be $k^*=5, 6, 5$ for ADVANCE, AHEAD-DS, and TAU 2019 respectively.

\subsection{Downstream Training}\label{sec:downstream_training}
To evaluate AuditoryHuM label efficacy, lightweight scene recognition models were trained on both baseline dataset provided and AuditoryHuM generated labels using $k^*$. The resulting mean average precision (mAP) and accuracy are reported in Table \ref{tab:downstream}. The same 70/10/20 train/validation/test splits were used for dataset provided and AuditoryHuM labels. The model was based on the OpenYAMNet\cite{zhong2026datasetmodelauditoryscene} pretrained on AudioSet\cite{gemmeke2017audio}. Full training details are in Table \ref{tab:training_hyperparams}.

Table \ref{tab:downstream} shows that AuditoryHuM clusters enhance learnability on the ADVANCE dataset, matching the baseline mean average precision (mAP) ($0.84$) while raising classification accuracy from $0.79$ to $0.84$. Conversely, accuracy scales down on AHEAD-DS ($0.98 \rightarrow 0.89$) and TAU 2019 ($0.95 \rightarrow 0.75$). This performance divergence directly correlates with acoustic scene complexity. While ADVANCE contains mostly isolated, clean sound events, AHEAD-DS and TAU 2019 contain dense, multilayered urban soundscapes. In these complex conditions, zero-shot models yield low, undifferentiated alignment scores across multiple target sources, causing acoustically similar samples to scatter across separate clusters.

Baseline labels manage to group acoustically similar, but noisy and multilayered sounds together because samples recorded in the same physical environment share location based metadata. The lower downstream accuracy of AuditoryHuM on complex datasets is a trade-off of rejecting location metadata in favor of a content driven taxonomy.
\begin{table*}
    \centering
    \caption{The mean average precision (mAP) and accuracy of the dataset provided and AuditoryHuM generated labels. Higher values are better.}
    \label{tab:downstream}
    \begin{tabularx}{\textwidth}{p{0.15\textwidth}p{0.16\textwidth}p{0.17\textwidth}p{0.13\textwidth}p{0.17\textwidth}}
    \hline
                 & Provided Labels mAP & Provided Labels Accuracy & AuditoryHuM Labels mAP & AuditoryHuM Labels Accuracy \\
    \hline
        ADVANCE  & 0.84 & 0.79 & 0.84 & 0.84 \\
        AHEAD-DS & 0.97 & 0.98 & 0.83 & 0.89 \\
        TAU 2019 & 0.94 & 0.95 & 0.76 & 0.75 \\
    \hline
    \end{tabularx}
\end{table*}

\subsection{Computation, Complexity, and Scalability}\label{sec:computation_complexity_and_scalability}
Table \ref{tab:complexity} outlines the complexity and multiply and accumulate computations (MACs) for each step of AuditoryHuM. The MACs were derived from an average of samples drawn from the test sets. The MLLM labelling phase was the most intensive, requiring almost 1T MACs per sample. Followed by computing the \emph{CLAP scores}, which requires almost 6.3G MACs per sample, and is repeated a quadratic number of times to the length of the dataset. This is followed by discovery of the optimal number of clusters, which has a cubic upper bound, and requires computing the silhouette score across the range $k=2$ to $k_{max}$.

The initial implementation achieved throughput roughly equal to the duration of the audio dataset, using a L4 GPU with the Hugging Face Transformers library. Subsequent evaluation using Llama.cpp on a less capable T4 GPU yielded at least a fivefold increase in throughput. Because all steps are parallelisable due to independent repetition across samples, the framework scales to larger datasets when deployed on distributed systems.

\subsection{Human Labelling Result}\label{sec:human_labelling_result}
\begin{figure*}
\centering
\captionsetup{justification=centering}
    \begin{subfigure}{0.49\textwidth}
        \includegraphics[width=\textwidth]{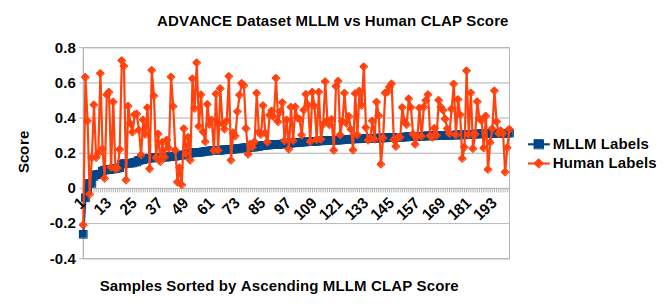}
        \caption{Comparison of \emph{CLAP scores} between human and MLLM labels. Sorted by ascending MLLM \emph{CLAP score} for the ADVANCE dataset.}
        \label{fig:human_strat1}
    \end{subfigure}
    \begin{subfigure}{0.49\textwidth}
        \includegraphics[width=\textwidth]{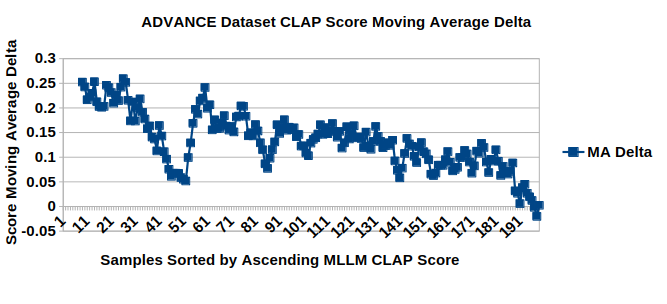}
        \caption{The moving average of the difference in \emph{CLAP scores} between human and MLLM labels. Window size 10, sorted by ascending MLLM \emph{CLAP score} for the ADVANCE dataset.}
        \label{fig:human_strat2}
    \end{subfigure}
    \caption{Comparison of \emph{CLAP scores} between human and MLLM labels.}
    \label{fig:human_strat}
\end{figure*}

The results in the previous sections demonstrate HITL provided labels are better aligned compared to the bottom $1\%$ performing cohort of MLLM labels. Further human labelling was performed with results shown in Figure \ref{fig:human_strat1} and \ref{fig:human_strat2}. The results illustrate alignment beginning to converge after manually annotating the bottom $4\%$ ($n = 200$) cohort of MLLM labels, for the ADVANCE dataset. This targeted intervention required only $1.1$ hours of human labour, an approximately $96.1\%$ reduction compared to the $28.2$ hours projected for full relabeling, while boosting alignment.

During evaluation, a single annotator conducted all human labelling tasks. Each 10 second sample required approximately 20 seconds of effort ($q \approx 20$ s), split between listening to the sample and annotating with a label. The annotator was provided with the MLLM label and its baseline \emph{CLAP score}. For this experiment, $R$ was unconstrained, the termination threshold was $s = 0$, and the smoothing window length was set to a heuristic value of $m = 10$.

\section{Discussion and Future Work}\label{sec:discussion_and_future_work}
The most sophisticated MLLM, Qwen 2.5 Omni 3B, produced the best-aligned labels but introduced hallucinations like non-English text and excessive length. Conversely, less advanced models avoided these errors but yielded lower alignment to audio content. Future work will focus on refining system prompts and introducing lightweight post generation filtering to eliminate structural artifacts without compromising alignment quality.

Strategically targeted HITL intervention proved invaluable for detecting unforeseen errors and guiding the design of the text cleanup pipeline, despite only improving the lowest 1\% of alignment scores. Analysis of low scoring cohorts highlighted a limitation of zero-shot techniques in complex, multilayered acoustic environments. This complexity caused acoustically similar, noisy samples to scatter across disparate clusters, resulting in a downstream accuracy drop for the AHEAD-DS and TAU 2019 datasets. Because this performance trade-off stems from rejecting location metadata in favor of a pure audio content taxonomy, future iterations will integrate source separation or explore multilabel zero-shot approaches with enhanced tolerance for overlapping soundscapes.

Finally, because the unsupervised clustering process lacks a definitive ground truth partition count ($k^*$), investigating task specific tuning heuristics remains a vital avenue. E.g. Binary applications like wind noise detection require a highly consolidated cluster count, whereas fine grained tasks discerning multiple environmental noises (wind, engine noise, etc) demand a higher cluster count. Developing adaptive, task aware thresholding techniques will maximise the utility of AuditoryHuM across diverse downstream applications.

\section{Conclusion}\label{sec:conclusion}
This paper introduced AuditoryHuM, a training-free framework that shifts the paradigm of audio dataset management from manual annotation to unsupervised, human-MLLM collaborative discovery. Our findings demonstrate that advanced MLLMs can successfully extract contextually accurate auditory scene labels that closely mirror human perception. Through strategic HITL intervention, we demonstrated an approximate 96.1\% reduction in human labor in the ADVANCE dataset, while mitigating hallucinations and refining edge cases. Clustering was then achieved using a human interpretable label alignment vector. Models trained on the clusters exhibit enhanced accuracy in clean acoustic environments compared to baseline (0.79 to 0.84), yet lower accuracy in dense, multilayered soundscapes due to limitations of zero-shot techniques. AuditoryHuM establishes a robust, training-free baseline for generating structured taxonomies for downstream auditory scene recognition models.

\begin{ack}
This work was supported by the Google DFI Catalyst fund and Macquarie University.
\end{ack}

\bibliographystyle{unsrt}
\bibliography{cluster_paper}

\newpage
\appendix
\section{Appendix}\label{sec:appendix}

\begin{table*}[h]
\centering
\caption{The prompts for generating labels.}
\label{tab:prompts}
    \begin{tabularx}{\textwidth}{p{0.7\textwidth}p{0.25\textwidth}}
        \hline
            Prompt & Notes \\
        \hline
            Describe the auditory scene using word pairs. Separate each pair with a comma. & Initial prompt. \\
        \hline
            Provide a short sentence to describe this set of audio samples. The frequency distribution of individual labels for this set of audio samples is provided: $<distribution> … </distribution>$. & Prompt to generate a descriptive composite label for the entire cluster. \\
        \hline
            Describe this auditory scene. & Captioning prompt.\\
        \hline
    \end{tabularx}
\end{table*}

\begin{algorithm}[h]
\caption{Cluster count selection via silhouette stability.}
\label{alg:cluster_selection}
\SetKwInOut{Input}{Input}\SetKwInOut{Output}{Output}

\Input{
    $s(k)$ for $k \in [2, k_{max}]$ \tcp*{Silhouette scores}
    $w$ \tcp*{Moving variance window size}
    $t$ \tcp*{Stability threshold, $0 \le t \le 1$}
}
\Output{$k^*$ \tcp*{Optimal number of clusters}}

\BlankLine
% Step 1: Compute moving variances
\For{$k \leftarrow 2$ \KwTo $k_{max} - w$}{
    $Var(k, w) \leftarrow \text{variance of } \{s(k), s(k+1), \dots, s(k+w-1)\}$\;
}

\BlankLine
% Step 2: Extract min and max variance
$Var_{min} \leftarrow \min_{2 \le k \le k_{max} - w} Var(k, w)$\;
$Var_{max} \max_{2 \le k \le k_{max} - w} Var(k, w)$\;

\BlankLine
% Step 3: Edge case handling
\If{$Var_{min} == Var_{max}$}{
    \Return \text{``No clustering should be performed''}\;
}

\BlankLine
% Step 4: Iterative threshold search
$k^* \leftarrow k_{max} - w$ \tcp*{Default if threshold is never breached}
\For{$k \leftarrow 2$ \KwTo $k_{max} - w$}{
    \If{$\frac{Var(k, w)}{Var_{max} - Var_{min}} \le t$}{
        $k^* \leftarrow k$\;
        \textbf{break}\; \tcp*{Stop at the first instance it no longer holds}
    }
}

\Return $k^*$\;
\end{algorithm}

\begin{table}[h]
    \centering
    \caption{Breakdown of the computational cost and complexity of each step in AuditoryHuM, using multiply and accumulate computations (MACs) and Big-O.}
    \label{tab:complexity}
    \begin{tabularx}{\textwidth}{p{0.05\textwidth}p{0.15\textwidth}p{0.1\textwidth}p{0.2\textwidth}p{0.4\textwidth}}
        \hline
            Step & Model or Algorithm & Params | $\approx$ MACs & Complexity & Notes \\
        \hline
            1 & Label generation (Qwen 2.5 Omni 3B) & 4.7035B | 971.995G & $O(H \cdot W \log W) + O(N \cdot P + N^2 \cdot D) + O(P + N \cdot D)$ & Steps include STFT, encoder, and decoder. $H$ is the number of hops. $W$ is the window length. $N$ is the total token length. $P$ number of params. $D$ is the hidden layer dimension. \\
            2 & CLAP score (Human-CLAP) & 153.507M | 6.287G & $O(H \cdot W \log W) + O(T_a^2 \cdot D + T_a \cdot D^2) + L \cdot O(1) + L \cdot O(T_t^2 \cdot D + T_t \cdot D^2) + L \cdot O(E)$ & Steps include STFT, audio encoder, label clean up on $L$ labels, text encoder on $L$ labels, and Cosine similarity on $L$ labels. $W$ is the window length. $T_a$ is the audio token length. $T_t$ is the text token length. $D$ is the hidden layer dimension. $E$ is the CLAP embedding length. \\
            3 & Select best aligned label & & $O(L)$ & Find the largest value in an unsorted sequence of length $L$. \\
              & & & & Steps $1$ to $3$ are repeated $A$ times. Where $A$ is the number of audio samples. \\
            4 & Create alignment vector (Human-CLAP) & 153.507M | 6.287G & $O(H \cdot W \log W) + O(T_a^2 \cdot D + T_a \cdot D^2) + L \cdot O(1) + L \cdot O(T_t^2 \cdot D + T_t \cdot D^2) + L \cdot O(E)$ & Steps include STFT, audio encoder, label clean up on $L$ labels, text encoder on $L$ labels, and Cosine similarity on $L$ labels. $W$ is the window length. $T_a$ is the audio token length. $T_t$ is the text token length. $D$ is the hidden layer dimension. $E$ is the CLAP embedding length. \\
              & & & & Step $4$ is repeated up to $A^2$ times to get alignment vectors. \\
            5 & Spectral Embedding & & $O(A^3 \cdot D)$ & $A$ is the number of audio samples. $D$ is the size of the feature. \\
            6 & K-Means & & $O(A \cdot D \cdot k \cdot i)$ & $A$ is the number of audio samples. $D$ is the size of the output from spectral embedding. $k$ is the number of clusters. $i$ is the number of iterations. \\
            7 & Silhouette Score & & $O(A^2 \cdot D)$ & $A$ is the number of audio samples. $D$ is the size of the alignment vector. \\
              & & & & Repeat steps $5, 6, 7$ for $k=2$ to $k_{max}$ to find the optimal number of clusters. \\
            8 & Generate composite label (Qwen 2.5 Omni 3B) & 4.7035B | 212.913G & $k^* \cdot O(N \cdot P + N^2 \cdot D) + O(P + N \cdot D)$ & $N$ is the total token length. This is repeated $k^*$ times, where $k^*$ is the optimal number of clusters. \\
        \hline
    \end{tabularx}
\end{table}

\begin{figure*}[h]
\centering
\captionsetup{justification=centering}
    \begin{subfigure}{0.45\textwidth}
        \includegraphics[width=\textwidth]{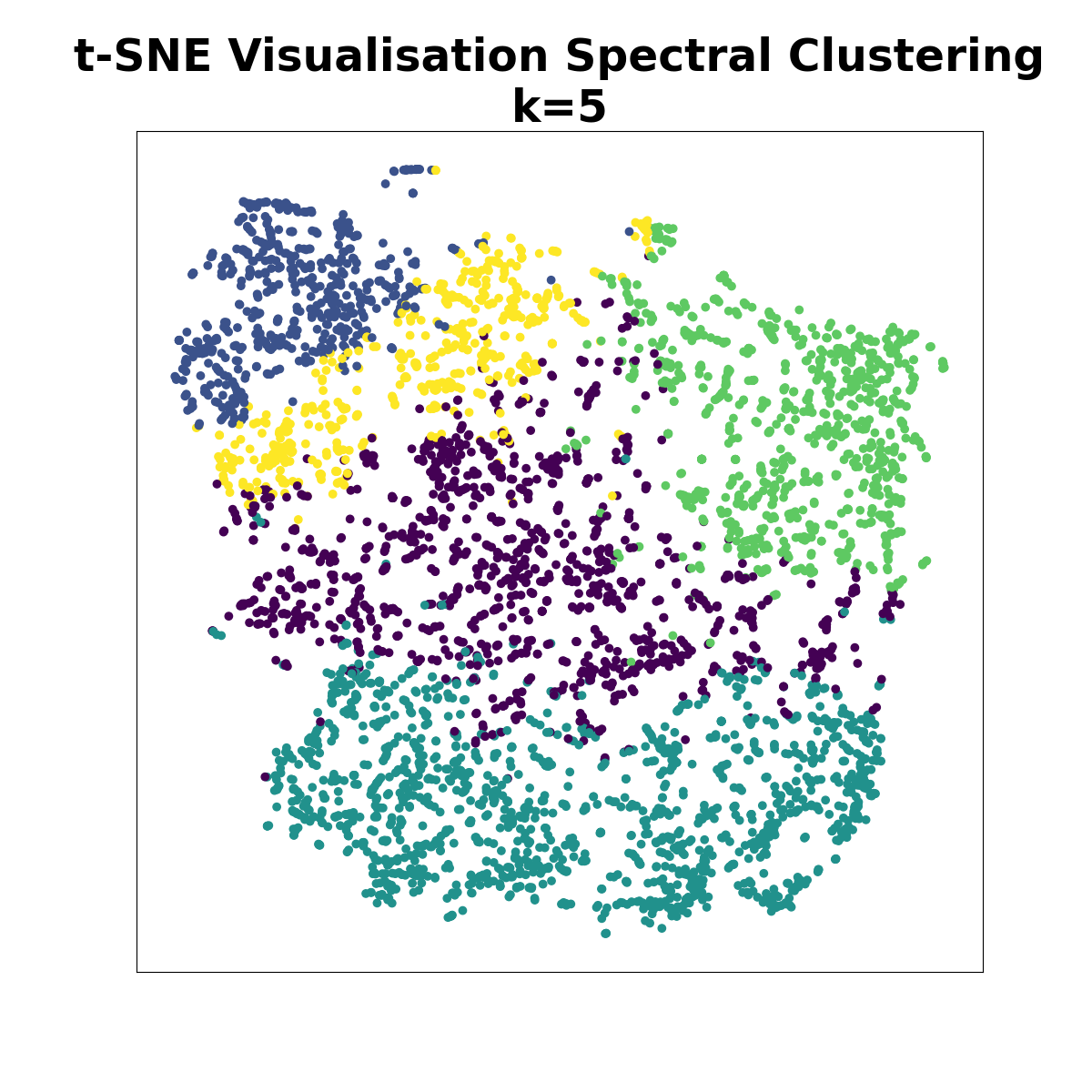}
        \caption{ADVANCE.}
    \end{subfigure}
    \begin{subfigure}{0.45\textwidth}
        \includegraphics[width=\textwidth]{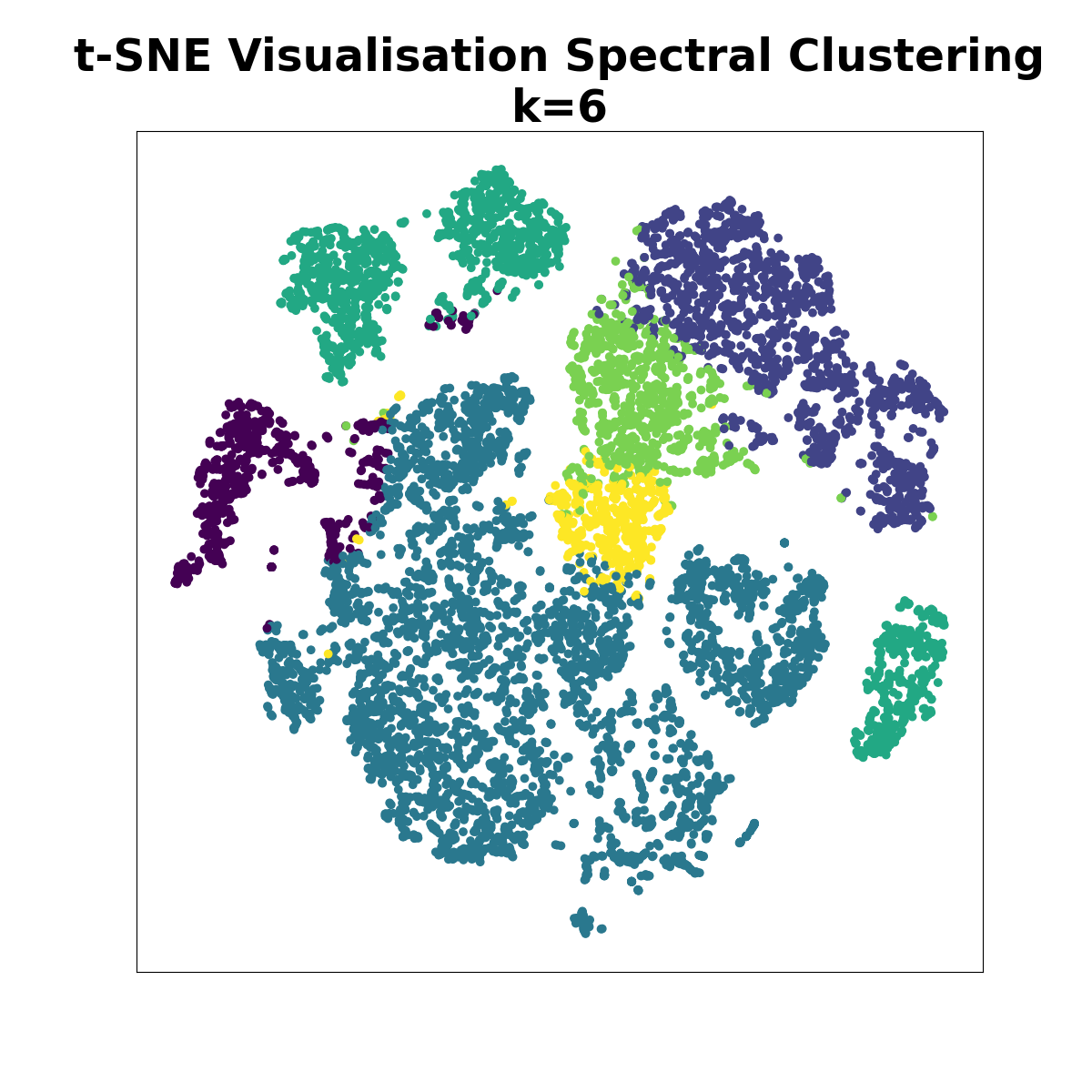}
        \caption{AHEAD-DS.}
    \end{subfigure}
    \begin{subfigure}{0.45\textwidth}
        \includegraphics[width=\textwidth]{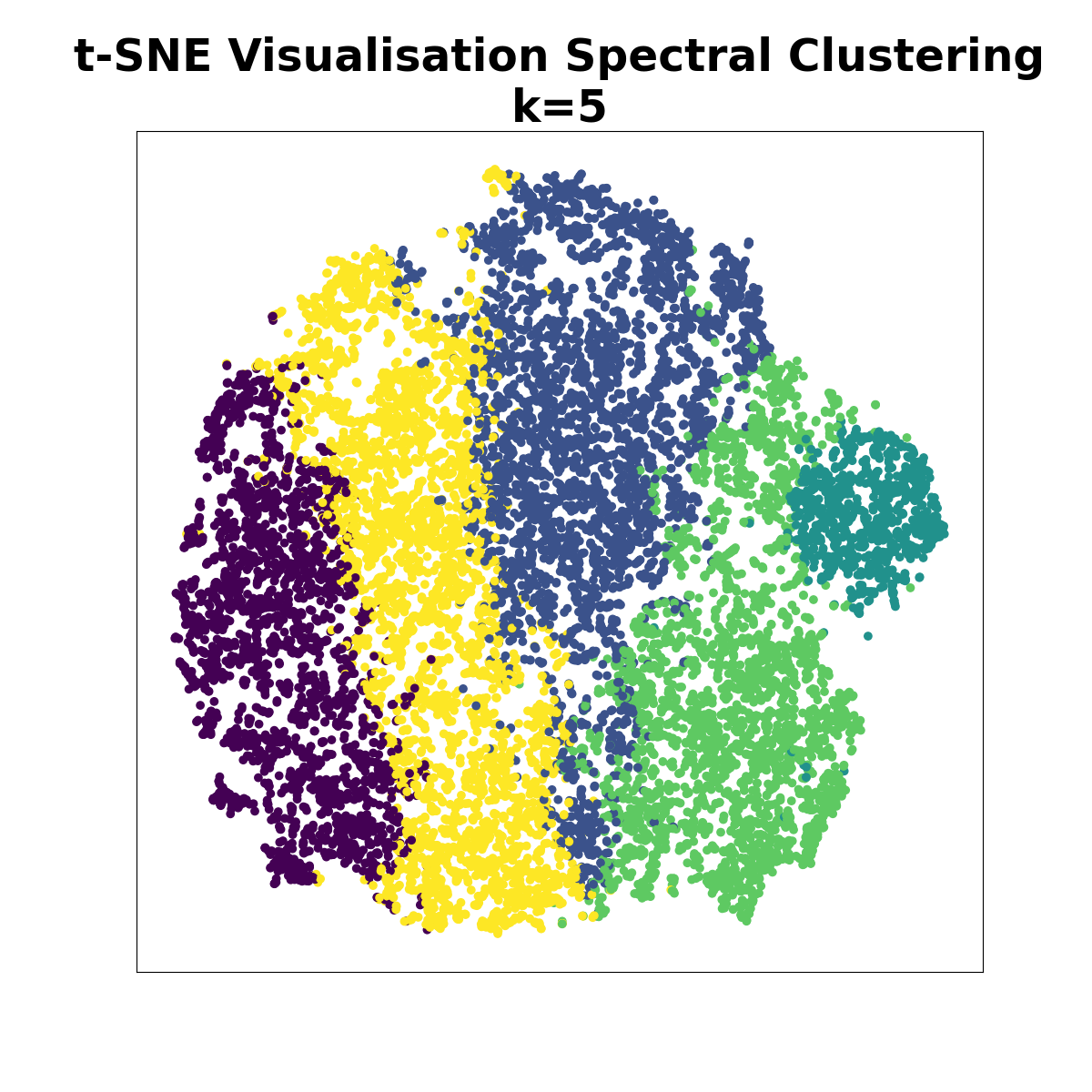}
        \caption{TAU 2019.}
    \end{subfigure}
    \caption{t-SNE visualisations using $k^*$ value for each dataset. Each unique colour represents a cluster of thematically related labels.}
    \label{fig:tsne}
\end{figure*}

\begin{table*}[h]
\centering
\caption{Details of the 3 largest clusters for each dataset.}
\label{tab:cluster_labels}
    \begin{tabularx}{\textwidth}{p{0.15\textwidth}p{0.15\textwidth}p{0.1\textwidth}p{0.5\textwidth}}
        \hline
            Dataset & Cluster Size / Total & Unique Labels & Composite Label for Cluster \\
        \hline
        ADVANCE  & 1582 / 5075  & 186 & The audio samples consist mainly of ocean related sounds such as waves, thunder, and wind, with a few other elements like car, train, and aircraft sounds mixed in, along with various background noises and other natural sounds. \\
        ADVANCE  & 1367 / 5075  & 375 & The audio samples consist of a variety of sounds including thuds, frying noises, car sounds, choir voices, vibrations, horse trotting, train movement, drills, sewing machines, and other mechanical sounds, among others. \\
        ADVANCE  &  983 / 5075  & 117 & These audio samples contain various animal sounds, human activities, and natural elements, including bird vocalizations, insects, and weather sounds. \\
        AHEADDS  & 4713 / 9968  & 198 & This audio set contains a variety of sounds including clatter, child speaking, restaurant background noise, white noise, engine sounds, female conversations, dishes clinking, door slams, car passing, male laughter, crowd talk, and many other common everyday sounds, totaling 4713 samples. \\
        AHEADDS  & 1668 / 9968  & 100 & The audio set contains a diverse range of music genres and styles, including country, blues, heavy metal, opera, jazz, and more, with a focus on vocal performances and specific instruments like guitar and piano. \\
        AHEADDS  & 1582 / 9968  &  93 & The audio samples predominantly feature traffic noise, car passing, and various road-related sounds, with some additional elements like engine sounds, wind, and weather effects. \\
        TAU 2019 & 3980 / 14400 & 331 & This set of audio samples includes various environmental sounds, such as bird noises, human activities, and traffic, covering a wide range of scenarios and activities. \\
        TAU 2019 & 3853 / 14400 & 298 & The audio set contains a diverse collection of sounds, including car sounds, conversations, traffic noises, and various human activities. \\
        TAU 2019 & 3053 / 14400 & 248 & The audio samples feature a variety of sounds, including rattling, car radio noise, air brake, honking, squeaking brakes, and more, from different contexts like vehicles, trains, and outdoor environments. \\
        \hline
     \end{tabularx}
\end{table*}

\begin{table*}[h]
\centering
\caption{Downstream training configuration and hyperparameters.}
\label{tab:training_hyperparams}
    \begin{tabularx}{\textwidth}{p{0.3\textwidth}p{0.65\textwidth}}
        \hline
        Component / Hyperparameter & Value / Strategy \\
        \hline
        Architecture & OpenYAMNet (Pretrained on AudioSet) \\
        Optimiser & Adam \\
        Loss Function & Focal Loss \\
        Label Smoothing & 0.1 \\
        Data Scheduling & Shuffled each epoch \\
        Learning Rate Scheduler & Halved after 3 consecutive epochs without val improvement \\
        Early Stopping & Terminated after 6 stagnant validation epochs \\
        Post processing & Each 10 second sample was divided into 20 windows of 960 milliseconds, with 480 millisecond overlaps. The results were collapsed by summing the 20 confidence values for each class, then a softmax was applied. \\
        \hline
    \end{tabularx}
\end{table*}

\newpage
\clearpage

\end{document}